\documentclass[12pt]{article}
\usepackage{graphicx}
\usepackage{hyperref}
\usepackage{dcolumn}
\usepackage{bm}

\begin{document}

\begin{center}
\LARGE{\bf{Has the QCD critical point been observed at RHIC?}}
\end{center}

\vspace*{1.7cm}
\begin{center}
N.~G.~Antoniou, N.~Davis and F.~K.~Diakonos\\
Department of Physics, University of Athens, GR-15784, Greece\\
\end{center}

\vspace*{2cm}

\begin{abstract}
The experimental search for the location of the QCD critical point in the phase diagram is of primary importance. In a recent publication it is claimed that measurements at RHIC lead not only to the location of the critical point ($\mu_{cep}=95$ MeV, $T_{cep}=165$ MeV) but also to the verification of its universality class ($3d$ Ising system) by extracting the values of the critical exponents ($\gamma=1.2$, $\nu=0.66$). We argue that this claim is based on an erroneous treatment of scaling relations near the critical point. As a result, the correct interpretation of the measurements cannot be linked to the QCD critical point.
\end{abstract}

\newpage

In a recent Letter \cite{Lacey2015} a search for the QCD critical  end point is performed by studying two-pion interferometry measurements in Au+Au ($\sqrt{s_{NN}}=7.7 - 200$ GeV) and Pb+Pb ($\sqrt{s_{NN}}=2.76$ TeV) collisions. The result is remarkable since the location and the universality class of the chiral critical point, are claimed to be extracted from the data ($\mu_b^{cep} \approx 95$ MeV, $T^{cep} \approx 165$ MeV, $\nu \approx 0.66$, $\gamma \approx 1.2$).

The basic argument in this work is that the difference $R_{out}^2 - R_{side}^2$, being a measure of the emission duration ($\Delta \tau$) \cite{Heinz1996,Pinkenburg1999} is directly linked to the compressibility according to the following relations
\begin{equation}
(R_{out}^2 - R_{side}^2) \sim (\Delta \tau)^2 \sim v_s^{-2} = \rho k_S
\label{eq1}
\end{equation}
where $v_s$ is the sound velocity, $\rho$ the density of the pion fluid and $k_S$ the adiabatic (isoentropic) compressibility. The assumption $R_{out}^2 - R_{side}^2 \sim k_S$ is therefore plausible and leads to a divergence of $k_S$ at the critical point ($k_S \to \infty$ when $T \to T_c$) following the vanishing of the sound velocity in the same limit. However the critical exponents which dictate the divergences in this approach, shown in the power-laws \cite{Stanley1971}
\begin{equation}
k_S \sim \vert T - T_c \vert^{-\alpha}~~;~~v_s \sim \vert T - T_c \vert^{\alpha/2}~~;~~\xi \sim \vert T - T_c \vert^{-\nu}
\label{eq2}
\end{equation}
lead to the finite-size scaling relation, in the author's notation
\begin{equation}
(R_{out}^2 - R_{side}^2)_{max} \sim \bar{R}^{\alpha/\nu}
\label{eq3}
\end{equation}
which is different from the equation (4) of the Letter, $(R_{out}^2 - R_{side}^2)_{max} \sim \bar{R}^{\gamma/\nu}$ since the exponents $\gamma$, $\alpha$ are interchanged. The reason is that the author has, contrary to his assumption and the plausibility argument (1), employed the isothermal compressibility, $k_T \sim \vert T - T_c \vert^{-\gamma}$, in the treatment, instead of the adiabatic one, $k_S$. As a consequence, the computed value of 
$\gamma$ in the publication is, in reality, the value of the exponent $\alpha$ and the solution for the claimed critical exponent is $\alpha \approx 1.15 \pm 0.065$ \cite{Stanley1971}. In fact $k_S$ is related to $k_T$ through the thermodynamic relation $k_S=k_T \frac{C_V}{C_p}$ leading to the scaling behaviour in Eq.~(\ref{eq2})
since
\begin{equation}
k_T \sim \vert T - T_c \vert^{-\gamma}~,~C_V \sim \vert T - T_c \vert^{-\alpha},~C_p \sim \vert T - T_c \vert^{-\gamma}
\label{eq4}
\end{equation} 
(see Refs. \cite{Stanley1971} and \cite{Huang1987}). Now since $\nu \approx 0.67 \pm 0.05$ the Josephson scaling relation $\nu d = 2 - \alpha$, for $d=3$ is strongly violated and the found solution for the pair $(\alpha,\nu)$ cannot represent any universality class in $3d$. Finally, even if one erroneously insists that the compressibility described by the observable $(R_{out}^2 - R_{side}^2)_{max}$ is the isothermal one, then in such a case it should be related, according to the fluctuation-dissipation theorem \cite{Huang1987}, to integrated fluctuations of the order parameter which in the case of the QCD critical point is not the pion density considered in the Letter but the density of the chiral condensate $\langle \bar{q} q \rangle$ or equivalently the net-baryon density. 

We conclude that in the Letter under discussion there is no self-consistent solution for the critical point. The most interesting experimental result, however, is the scaling property illustrated in figure 5 of the Letter which may indicate a new phenomenon in the crossover regime, associated with the hadronization process at small values of the chemical potential. This experimental observation is worth studying further and in depth, whereas the finite size scaling relation (4) in the publication (but with the exponent $\alpha$ instead of $\gamma$) remains a powerful tool in the search for the critical point at higher values of the chemical potential. \\

\end{document}